\documentclass[reprint, superscriptaddress, amsmath, amssymb, aps, pra, a4]{revtex4-2}
\usepackage{graphicx}
\usepackage{dcolumn}
\usepackage{hyperref}
\usepackage{braket}

\begin{document}

\title{BBM92 quantum key distribution over a free space dusty channel of 200 meters}

\author{Sarika Mishra}
\affiliation{Quantum Science and Technology Laboratory, Physical Research Laboratory, Ahmedabad, India 380009.}
\affiliation{Indian Institute of Technology, Gandhinagar, India 382355.}

\author{Ayan Biswas}
\affiliation{Quantum Science and Technology Laboratory, Physical Research Laboratory, Ahmedabad, India 380009.}
\affiliation{Indian Institute of Technology, Gandhinagar, India 382355.}

\author{Satyajeet Patil}
\affiliation{Quantum Science and Technology Laboratory, Physical Research Laboratory, Ahmedabad, India 380009.}
\affiliation{Indian Institute of Technology, Gandhinagar, India 382355.}

\author{Pooja Chandravanshi}
\affiliation{Quantum Science and Technology Laboratory, Physical Research Laboratory, Ahmedabad, India 380009.}

\author{Vardaan Mongia}
\affiliation{Quantum Science and Technology Laboratory, Physical Research Laboratory, Ahmedabad, India 380009.}
\affiliation{Indian Institute of Technology, Gandhinagar, India 382355.}

\author{Tanya Sharma}
\affiliation{Quantum Science and Technology Laboratory, Physical Research Laboratory, Ahmedabad, India 380009.}
\affiliation{Indian Institute of Technology, Gandhinagar, India 382355.}

\author{Anju Rani}
\affiliation{Quantum Science and Technology Laboratory, Physical Research Laboratory, Ahmedabad, India 380009.}
\affiliation{Indian Institute of Technology, Gandhinagar, India 382355.}

\author{Shashi Prabhakar}
\affiliation{Quantum Science and Technology Laboratory, Physical Research Laboratory, Ahmedabad, India 380009.}

\author{S. Ramachandran}
\affiliation{Aerosol Monitoring Laboratory, Physical Research Laboratory, Ahmedabad, India 380009.}

\author{Ravindra P. Singh} \email{rpsingh@prl.res.in}
\affiliation{Quantum Science and Technology Laboratory, Physical Research Laboratory, Ahmedabad, India 380009.}

\date{\today}

\begin{abstract}
Free space quantum communication assumes importance as it is a precursor for satellite-based quantum communication needed for secure key distribution over longer distances. 
Prepare and measure protocols like BB84 consider the satellite as a trusted device, which is fraught with security threat looking at the current trend for satellite-based optical communication. 
Therefore, entanglement-based protocols must be preferred, so that one can consider the satellite as an untrusted device too. 
The current work reports the implementation of BBM92 protocol, an entanglement-based QKD protocol over 200 m distance using an indigenous facility developed at Physical Research Laboratory (PRL), Ahmedabad, India. 
Our results show the effect of atmospheric aerosols on sift key rate, and eventually, secure key rate. 
Such experiments are important to validate the models to account for the atmospheric effects on the key rates achieved through satellite-based QKD.
\end{abstract}

\maketitle

\section{\label{sec:intro}Introduction}

In classical communication, the security of encryption keys for parties communicating with each other is an ongoing challenge. 
Even after 50 years of digital communication, the security depends upon the hardness of breaking the encryption, which may compromise the security of encrypted messages sent through a public channel once a quantum computer intercepts them. 
With the development of quantum computers having sufficient numbers of good quality qubits becoming a practical reality, the demand for secure communication has increased. 
It has already been realized that by using Shor's algorithm \cite{shor1994algorithms, beckman1996efficient, ekert1996quantum}, one can break the encryption used in key distribution between communicating parties \cite{gisin2002quantum}. 
Quantum Key Distribution (QKD), on the other hand, relies on the principles of quantum mechanics, like uncertainty principle, no-cloning theorem and monogamy of entanglement to securely distribute keys between the communicating parties \cite{bennett1984g, ekert1991quantum}. 
These principles make QKD safe even against a quantum computer while making no assumptions on Eve's technological capability.

Every protocol has distance limitations as the loss and disturbance in the channel increases with the transmission distance. 
Entanglement based QKD (EBQKD) are ideal for long-distance quantum communication like satellite-to-ground, as it can make two far apart ground stations communicate securely.
Also, the security is not compromised irrespective of the satellite distributing entangled photon pairs is trusted or not. 
It must be noted that the satellite-to-ground EBQKD has already been performed \cite{PhysRevLett.120.030501, PERUMANGATT20171858}.
The current limitation of using EBQKD is the low key rate, as mostly the entangled photon pairs produced are from spontaneous parametric down-conversion (SPDC) \cite{PhysRevLett.18.732, PhysRevLett.75.4337} process that is not very efficient. 
The two factors contributing to low key rates are low-efficiency of the photon-pair generation, and loss of photons in the communication channel. 
Low efficiency can be compensated with the high-brightness high-fidelity photon-pair sources. 
However, one of the ways to mitigate the effect of channel is by studying its effect on parameters controlling the QKD.
In the past, the controlled and quantitative studies to observe the influence of atmospheric conditions such as the presence of particulate matter (PM) or aerosols have never been considered.
In this study, we account the presence of aerosols and show its effect on the secure key rate. 

Atmospheric aerosols, solid or liquid particles suspended in air, are produced by natural sources and anthropogenic emissions. Mineral dust and sea salt are produced from natural sources while sulfate, nitrate, black carbon and organic carbon are emitted primarily and/or converted into particles through a gas-to-particle conversion mechanism. 
Atmospheric aerosols/PM can scatter and absorb the incident radiation.  
The dry and long summer of Ahmedabad, a metropolitan city and urban environment, contains a relatively higher amount of aerosols \cite{rajesh2020extensive}. 
For carrying out satellite-to-ground based communication, entanglement based QKD (EBQKD), such as BBM92 protocol is the most suitable as it does not require a trusted node \cite{PhysRevA.65.052310, Scherer:11}.
In this article, we report the demonstration of BBM92 protocol at the indigenously developed communication channel facility at PRL, and the effect of dust/atmospheric channel on the secure key rate. 
This is the first attempt, in India, to demonstrate such studies, and eventually will form a strong base for the future long-distance quantum communication.

The article is structured as follows. 
The section \ref{sec:back} contains the background information about the BBM92 protocol, the entangled photon-pair source, and the indigenous facility built to study the atmospheric channel. 
The section \ref{sec:result} contains the results obtained and the related discussion. And finally, we conclude in section \ref{sec:conc}.

\section{\label{sec:back}Background}
\begin{figure}[h]
    \centering
    \includegraphics[width=8cm]{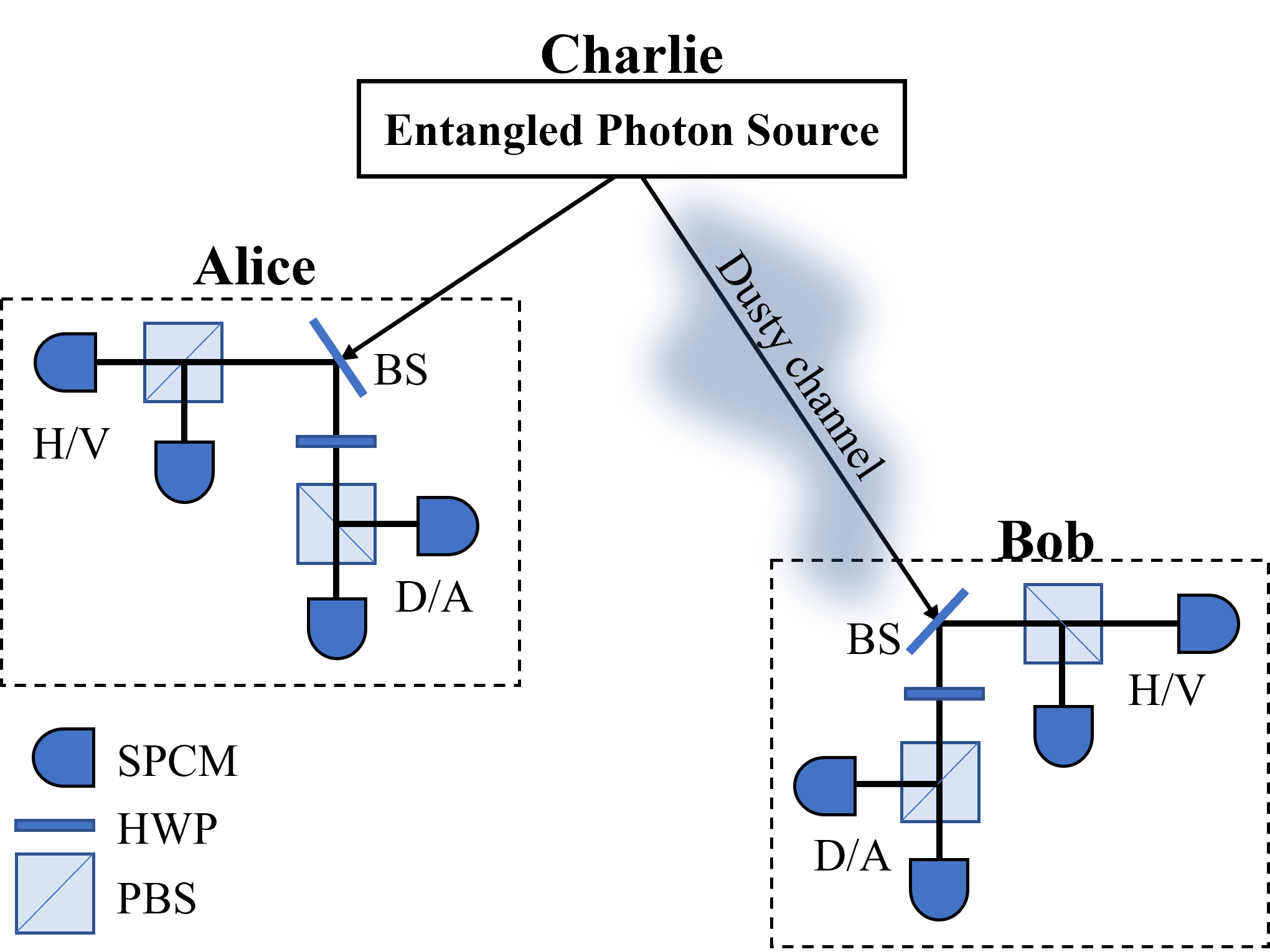}
    \caption{The block diagram of BBM92 protocol.}
    \label{fig:protocol}
\end{figure}

BBM92 is a QKD protocol which involves pairs of entangled photons and can be regarded as an entanglement-based version of the BB84 protocol. 
BB84 is a prepare and measure based QKD protocol where Alice randomly generates polarization states using RNGs whereas in BBM92 the randomness is inherent in the measurement of the entangled photon pairs. 
The basic block diagram of the BBM92 protocol is shown in Fig \ref{fig:protocol}.

In this protocol, a common sender Charlie generates a pair of entangled photons and sends them to Alice and Bob through a quantum channel. 
Quantum channels can be free space, water or optical fibre. 
Alice and Bob independently perform their measurements in a random basis. 
Once the measurement is done, both declare their basis choices through the public channel. 
Only those measurements contribute to key for which Alice and Bob have chosen the same basis, and the rest of the measurements are discarded. 
The key formed after this process is called sifted key. 
In addition, error correction (EC) and privacy amplification (PA) is performed on sifted key to get a secure key. 
In our experiment, the quantum channel between Charlie and Alice is not exposed to atmospheric aerosols as both are co-located in the same room. 
However, the quantum channel between Charlie and Bob experiences the free-space dusty atmospheric channel of 35 m and 200 m.

\subsection{\label{subsec:pairsource}Entangled photon-pair source}
\begin{figure}[h]
    \centering
    \includegraphics[width=9cm]{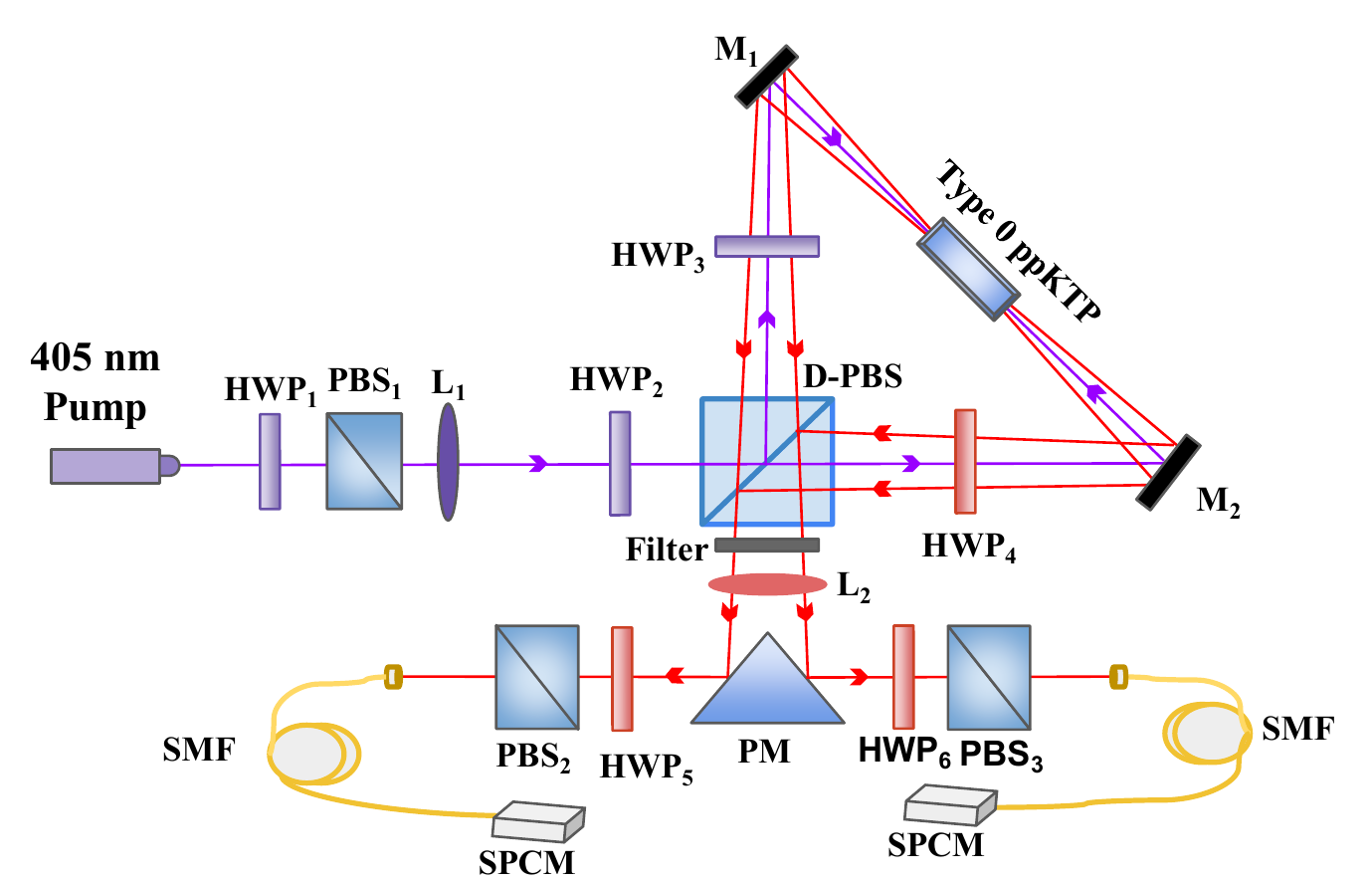}
    \caption{Schematic diagram of the polarization entangled photon pairs source. D-PBS operates at both 405 nm and 810 nm. HWP$_{1,2,3}$ and HWP$_{4,5,6}$ operate at 405 nm and 810 nm respectively. Lens L$_1$=400 mm, L$_2$=200 mm, M$_1$ and M$_2$: high reflecting mirrors at 810 nm, Filter: interference bandpass filter of bandwidth 10 nm centered at 810 nm, PM: prism mirror, SMF: Single-mode fiber, SPCM: single-photon counting modules (Excelitas AQRH-16-FC).}
    \label{fig:polsource}
\end{figure}

The high-efficient, high-brightness, in-house developed polarization-entangled photon-pair source was set up using ppKTP crystal placed in a Sagnac interferometer. 
The schematic of the experimental setup is shown in Fig. \ref{fig:polsource}\cite{jabir2017robust}. 
A continuous-wave laser at the wavelength of 405 nm and output power of $\sim$5 mW is used to pump a 30 mm long type-0 ppKTP crystal of period $\Lambda$ = 3.425 $\mu$m. 
A lens $L_1$ of focal length 400 mm is used to focus the pump beam on the crystal to generate entangled photons using a novel experimental scheme based on polarization Sagnac interferometer consisting of a dual-wavelength polarizing beam splitter cube (D-PBS), two half-wave plates (HWP) HWP$_3$, HWP$_4$, and two high reflecting (R$>$99\%) mirrors M$_1$, M$_2$ at 810 nm. 
The working principle of this setup is well explained in the article \cite{jabir2017robust}. 

Since both the clockwise and counter-clockwise pump beams follow the same path in opposite directions inside Sagnac interferometer and the ppKTP crystal is placed symmetric to the D-PBS, the implemented scheme is robust against any optical path changes to produce SPDC photons in orthogonal polarizations with ultra-stable phase. 
The analyzer comprises a PBS and a HWP plate, and is used to measure the polarization entanglement of the generated photon pairs. 
The polarization entangled state generated from this method is
\begin{equation}
    \ket{\psi} =\frac{1}{\sqrt{2}} \left( \ket{HH}+\ket{VV} \right).
\end{equation}

\subsection{\label{subsec:comm_channel}Communication channel and their arrangement}
\begin{figure*}
    \centering
   \includegraphics[width=\textwidth]{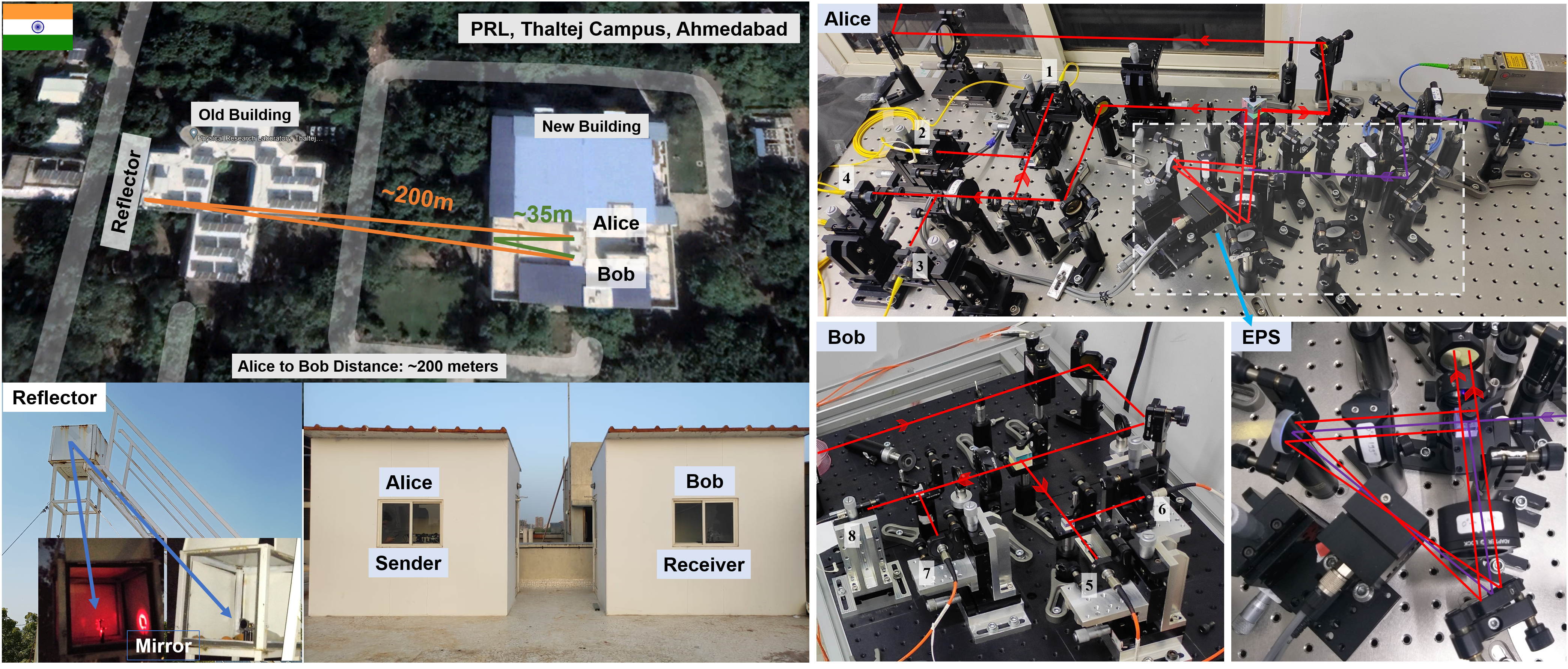}
    \caption{Arrangement of various components in the channel. It includes the location of Alice and Bob, and their setups.}
    \label{fig:channel}
\end{figure*}

\begin{figure*}
    \centering
    \includegraphics[width=\textwidth]{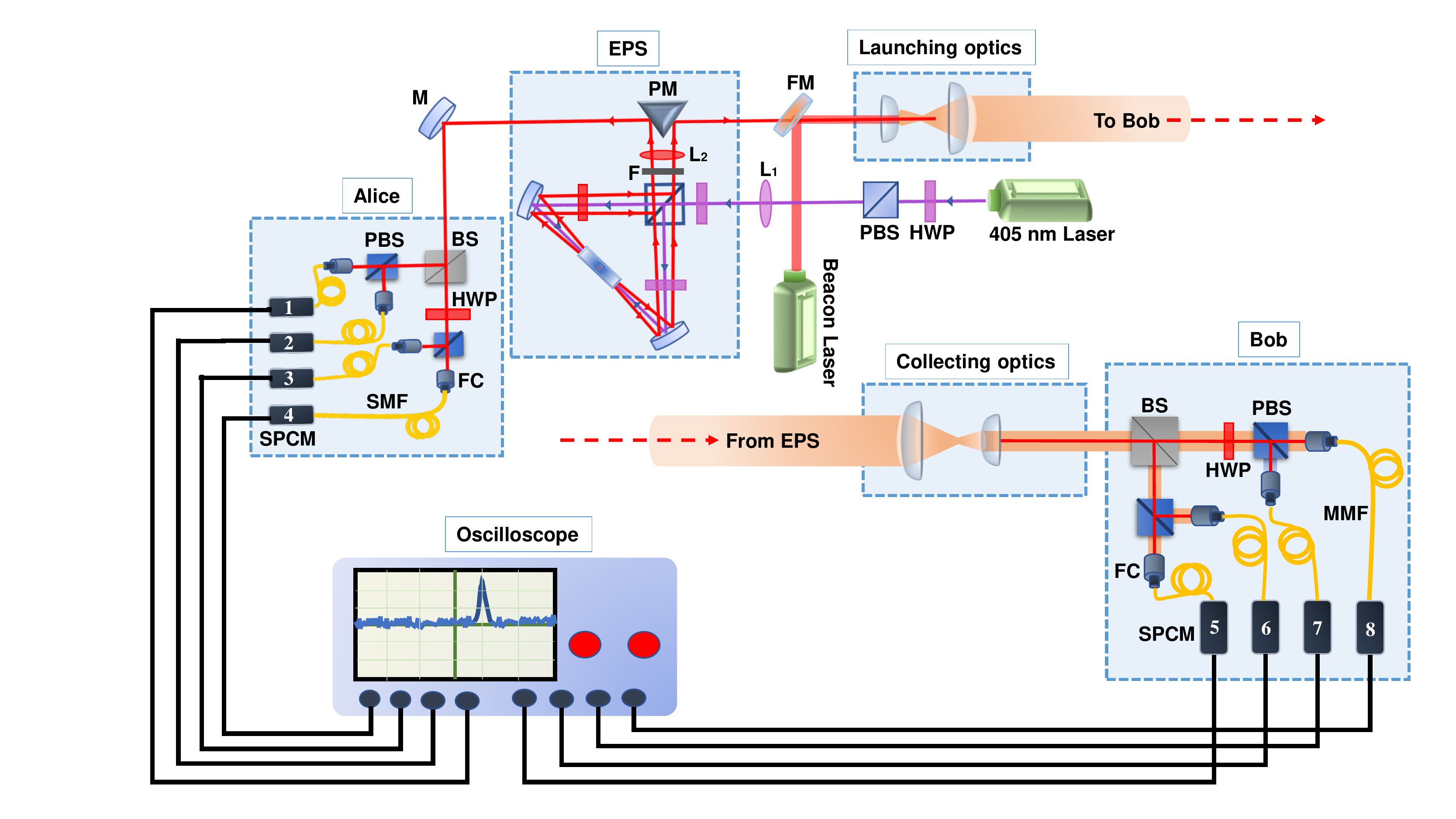}
    \caption{Schematic of the complete experimental setup is shown. This includes both optical and electronic arrangements. EPS: entangled photon source, FM: flip mirror, PM: prism mirror, M: mirror, F: filter, FC: fiber coupler, BS: beam splitter, PBS: polarization beam splitter, HWP: half wave plate, SMF: single-mode fiber, MMF: multi-mode fiber, SPCM: single-photon counting modules.}
    \label{fig:schematic}
\end{figure*}

Our communication channel consists of arrangements on two buildings, located nearby, in the Thaltej campus of PRL. 
The sending and receiving ends are located on the same terrace in two separate rooms. 
There are two reflectors - one for a path length of 35 m and another for 200 m on the other building to reflect back the photons.
The reflector consists of a 1-inch diameter mirror placed on a structure dedicated to this purpose. 
The channels run through a free-space open atmosphere to study the effect of aerosols/dust on the protocols. 
The schematic of the building structure and the placement of the equipment is shown in Fig. \ref{fig:channel}. 
Charlie who prepares entangled photon-pairs and Alice are co-located in our arrangement, while Bob is placed in the nearby room. 
Out of each pair generated, one stays at the receiving end with Alice and the second photon travels through the quantum channel consisting of going to the reflector and getting reflected to reach the receiving end to Bob, as shown in the protocol diagram (Fig. \ref{fig:protocol}).
The complete optical setup is shown in Fig. \ref{fig:schematic}. 
The experiment was performed at night (11 PM to 5 AM Indian Standard Time), in order to ensure that there is no interference due to the direct sunlight.

\section{\label{sec:result}Results and Discussion}
To study the effect of aerosols/dust present in the atmospheric channel on the secure key rate, we performed the experiment for 35 m channel on 8 May 2021 and 200 m channel on 10 May 2021. 
The atmospheric conditions for these two days are summarized in Table \ref{tab:atmoscond}. 
The extinction coefficient (Ext.) in the table represents the loss of photons due to absorption and scattering, which depends on the concentration of atmospheric aerosols, their size distribution, and refractive index. 
Aerosol characteristics reported here are measured in Aerosol Monitoring Laboratory, PRL. 
The extinction coefficient is measured using CAPS PM monitor (Aerodyne Research Inc., USA) and aerosol size distribution and PM less than 2.5 micrometer diameter are measured using an aerosol spectrometer (GRIMM Aerosol Technik, Germany).
From the table, the extinction coefficient on 10 May 2021 is lower by $\sim$37\%, leading to clear atmosphere and more transmissivity through channel. 
Also, the value of particulate matter (PM2.5) has lower concentration. 
Thus, choosing these two different days have provided the variability in the atmospheric conditions, most suitable for this study. 
In Ahmedabad, the higher values occur due to lower wind speed and shallow atmospheric boundary layer \cite{rajesh2020extensive}. 

\begin{table}[h]
    \centering
    \begin{tabular}{|l|l|l|}
        \hline 
        \textbf{Date} & \textbf{Ext. (Mm$^{-1}$)} & \textbf{PM2.5 ($\mu$g/m$^3$)} \\ 
        \hline 
        8 May 2021 & $\gamma_{35}=76.41\pm$7.78 & 2.87$\pm$0.26 \\ 
        \hline 
        10 May 2021 & $\gamma_{200}=48.67\pm$6.70 & 1.68$\pm$0.24 \\ 
        \hline 
    \end{tabular} 
    \caption{Extinction coefficient (Ext.) and particulate matter (PM2.5) concentrations are averages of hourly data from 12 midnight to 5 AM obtained from the Aerosol Monitoring Laboratory, PRL. The extinction coefficients data correspond to 525 nm. The data is fitted with the Angstrom power law for urban aerosols (model) corresponding to 70\% relative humidity (appropriate for April-May) and found that wavelength exponent for extinction coefficients between 525 and 800 nm is about 1.5.}
    \label{tab:atmoscond}
\end{table}

The channel transmissivity was measured by sending the output from a beacon laser of 810 nm through the channel and was measured to be 94\% for 35 m and 70\% for 200 m. 
The power of laser was measured before the launching optics and after the collecting optics to estimate the channel transmissivity on both days. 
The use of a beacon laser was crucial for precise alignment and also for further corrections required due to the breeze in the weather on both days.

From Fig. \ref{fig:schematic}, one can observe that eight detectors are used in the experiment, four each for Alice and Bob. 
Analogous to any classical communication protocol, BBM92 also requires timing synchronization between communicating parties to distribute keys correctly. 
We connected the SPCM4 and SPCM8 to time-tagger ID900, and found the time-difference between Alice and Bob is 120 ns for 35 m and 666 ns for 200 m channels based on the histogram obtained between the two channels. 
The coincidences were measured with the time-window of 1 ns.

\begin{figure}[h]
    \centering
    \includegraphics[width=8cm]{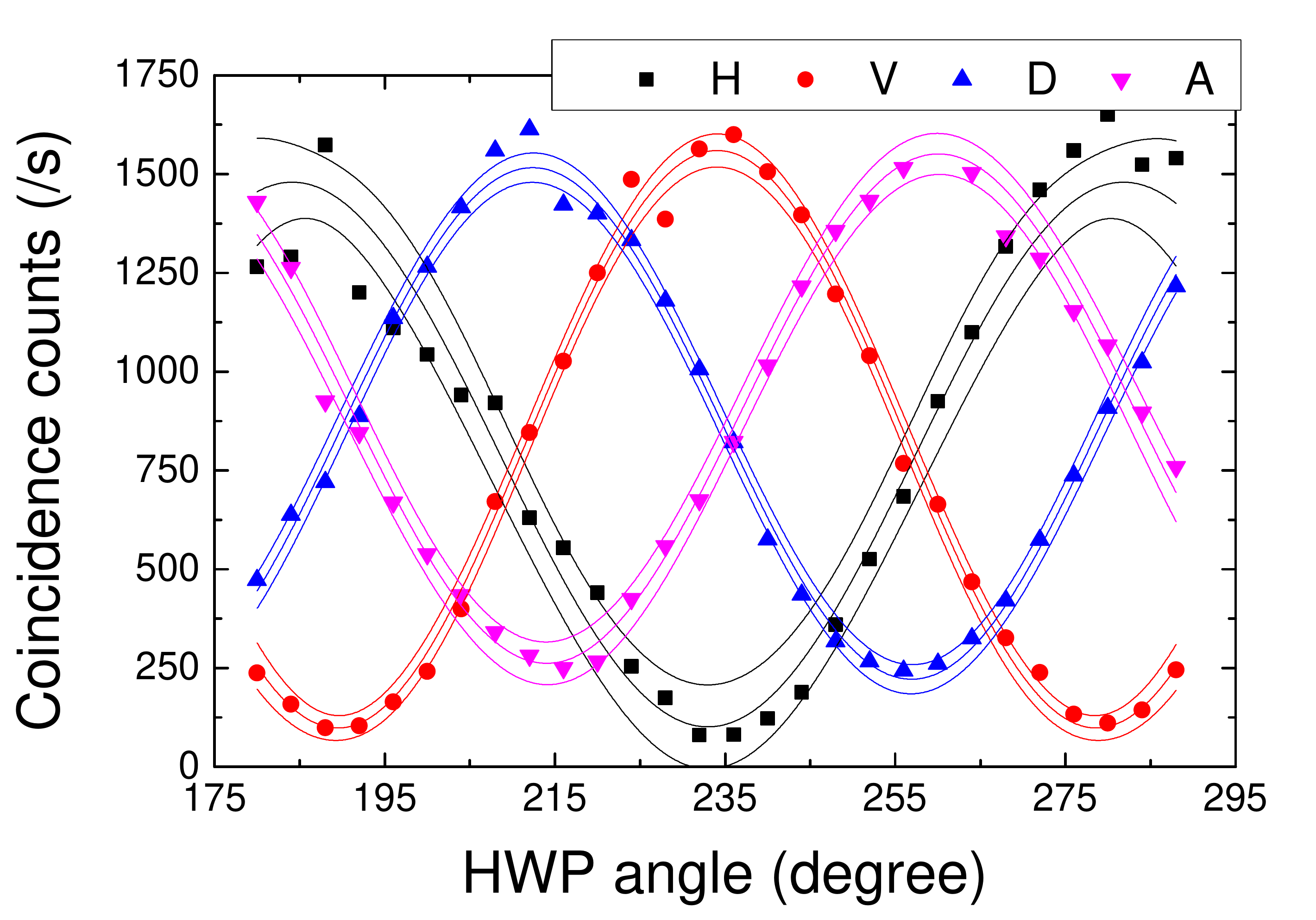}
    \caption{Obtained H/V/D/A graphs for 35 m channel taken on 8 May 2021. Solid lines are theoretical fit and lines of 98\% confidence interval.}
    \label{fig:results35m}
\end{figure}

\begin{figure}[h]
    \centering
    \includegraphics[width=8cm]{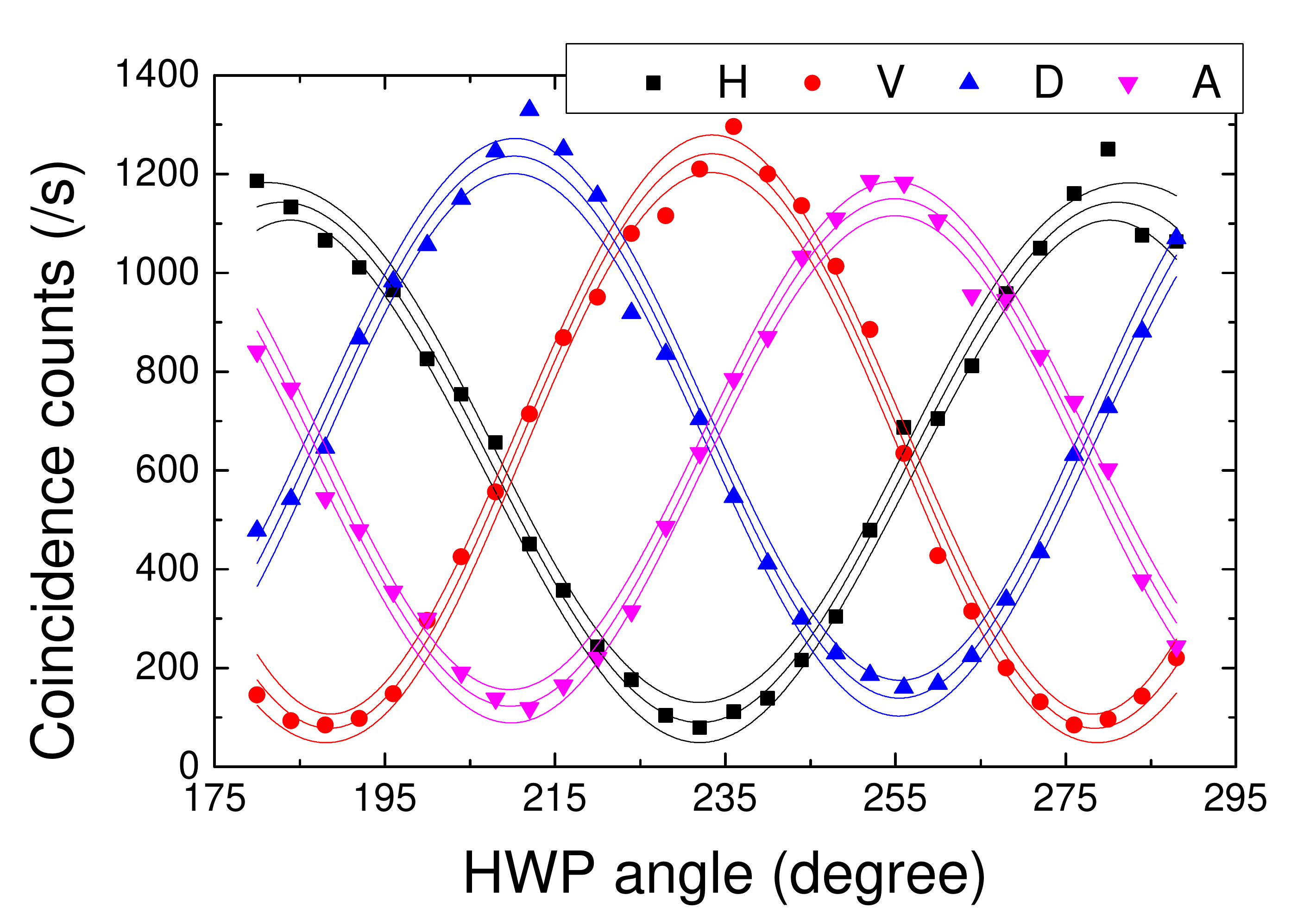}
    \caption{Obtained H/V/D/A graphs for 200 m channel taken on 10 May 2021. Solid lines are theoretical fit and lines of 98\% confidence interval.}
    \label{fig:results200m}
\end{figure}

We started with the channel effect on the entanglement distribution between Alice and Bob, and measured the H/V/D/A polarisation visibilities.  
The graphs corresponding to these measurements are shown in Figs. \ref{fig:results35m}-\ref{fig:results200m}. 
The visibilities obtained for 35 m channel on 8 May 2021 were 93.17\%, 93.71\%, 85.39\% and 83.12\% for H/V/D/A, respectively. The mean of these visibilities is 88.85\%.
For 200 m channel on 10 May 2021, the respective visibilities were 92.16\%, 93.72\%, 88.76\% and 89.34\%. The mean of these visibilities is 90.99\%.
The parameters obtained for both the distances are summarized in Table \ref{table:result}.
The $S$ value greater than $2$ ensures that secure key distribution can be done over the given channel. 
It shows that the atmosphere plays a role as visibilities are higher for 200 m, because of the lower extinction coefficients on May 10, even if the channel length is longer. 
The total count rate is a combined effect of channel losses and the length of the channel.

The quantum bit-error-rate ($Q$) and sifted key rate (SKR) can be calculated as \cite{scheidl2009feasibility, PhysRevA.104.022406}
\begin{eqnarray}
    Q&=&\frac{1-Visibility}{2}.
\end{eqnarray}
The value of $Q$ for 35 m is 5.58\% and for 200 m it is 4.50\%. The expression for sifted key rate can be calculated from the coincidence counts rates ($cc$) as
\begin{eqnarray}
    R_{sif}&=& 4 \times cc.
\end{eqnarray}
The factor $4$ accounts for the total of all four coincidences ($HH$, $VV$, $DD$ and $AA$), as the correct number of coincidences will contribute to total sifted key rate. 
The secret key rate after the error correction is given by
\begin{eqnarray}
    R_{sec}&=& n \left(\frac{1}{2}-2 \times Q -s \right),
\end{eqnarray}
where $n$ is the number of bits left after error correction, and $s$ is the security parameter \cite{476316}. 
The lower value of $R_{sec}$ is a direct effect of worse atmospheric conditions, which is also consistent with the visibilities. 
The different key rates obtained for both the distances are summarized in Table \ref{table:result}.

\begin{table}[h]
    \centering
    \begin{tabular}{|l|l|l|}
        \hline 
        \textbf{Parameters} & \textbf{35 m} & \textbf{200 m} \\ 
        \hline
        \textbf{Channel transmission (\%)} & 94 & 70 \\
        \hline
        \textbf{CHSH Bell parameter (S)} & 2.51$\pm$0.06 & 2.54$\pm$0.06 \\
        \hline
        \textbf{Mean visibility (\%)} & 88.85$\pm$5.39 & 90.99$\pm$5.89 \\
        \hline 
        \textbf{QBER (\%)} & 5.58 & 4.50 \\ 
        \hline 
        \textbf{Sifted key rate (kbps)} & 6.37 & 4.89 \\ 
        \hline 
        \textbf{Key rate after EC (kbps)} & 6.01 & 4.20 \\ 
        \hline 
        \textbf{Key rate after PA (kbps)} & 2.33 & 1.71 \\ 
        \hline 
        \textbf{Secure key rate (kbps)} & 2.33 & 1.71 \\ 
        \hline 
    \end{tabular} 
    \caption{Summary of the parameters obtained for 35 m channel on 8 May 2021, and 200 m channel on 10 May 2021.}
    \label{table:result}
\end{table}

The measured channel transmission is 0.94 for 35 m and 0.70 for 200 m on the two dates of the experiment. 
As the atmospheric condition follows Beer-Lambert law \cite{beer1852bestimmung}, the transmission $T$ of the channel can be written in terms of the number of photons sent ($N_{In}$) and the number of photons received ($N_{Out}$) on the other end of the free space atmospheric channel as 
\begin{equation} \label{eq:BLlaw}
    T=\frac{N_{Out}}{N_{In}}=Sc~\exp (-1.5\times\gamma L),
\end{equation}
where $Sc$ is the scaling parameter, $\gamma$ is the extinction coefficient of the atmosphere, and $L$ is the propagation length. 
The scaling parameter contains all the losses, except the effect from atmosphere. With the transmission information, we obtained the values $Sc_{35}=0.944$ and $Sc_{200}=0.710$. 

Comparing the atmospheric conditions including the channel lengths for 8 May and 10 May, the $T$ ratios  can be written as
\begin{equation}\label{eq:factor}
    f=\frac{Sc_{35}~\exp (-1.5\times\gamma_{35} 35)}{Sc_{200}~\exp (-1.5\times\gamma_{200} 200)}=1.34.
\end{equation}
From our channel and experiments, the similar ratios are observed for the coincidence rates ($cc$), sifted key rate ($R_{sif}$) and secured key rates ($R_{sec}$), as shown by 
\begin{eqnarray}\label{eq:calc}
    \frac{cc_{35}}{cc_{200}}&=&1.28, \\
    \frac{R_{sif~35}}{R_{sif~200}}&=&1.30, {~ ~ \rm and} \\
    \frac{R_{sec~35}}{R_{sec~200}}&=&1.36.
\end{eqnarray}
The similar values show that the Beer-Lambert law is suitable for considering the atmospheric aerosols and is consistent with the quantum parameters observed from the channel.
One of the reasons is the linear dependence of sifted key rate on the coincidences observed per second. 
Most of the parameters like visibility, QBER and CHSH Bell parameter do not depend on the atmospheric channel, as the channel does not introduce polarization changes to the photons. 
To obtain the key rates for any other day, the new factor $f$ need to be calculated from Eq. \ref{eq:factor} and similarly the other parameters can be obtained using Eqs \ref{eq:calc}.

To determine the secure key, the time tags for all the eight channels have been recorded using the eight-channel digital oscilloscope (Tektronix MSO68B 6-BW-2500).
The oscilloscope has a bandwidth of 2.5 GHz, and it was used to record the TTL pulses coming from single-photon detectors with a sampling rate of 1.25 GS/s with an integration time of 40 ms to measure the key rate. 
The recorded data contains the voltage sample, and from that, the rising edge of TTL pulses provides the arrival time.
The post-processing includes basis reconciliation, error estimation, error correction, and privacy amplification, which are implemented in MATLAB. 
We have implemented a low-density parity-check (LDPC) error correction code \cite{limei2020qkd} with a code rate of 0.5 for error correction. 
Choice of LDPC matrices closely depends on QBER and directly affects the secure key rate. Higher QBER means it takes more parity bits to correct those errors and thus, reduces the key rate. 

Once error-corrected keys are generated, privacy amplification \cite{GPA} is done via 2-universal Toeplitz hash function.
During privacy amplification, the length of the hashed output is chosen pertaining to the parity bits used in error correction, QBER, and security parameter which decides the final length of the secure key.
The data analysis/post-processing of entanglement-based protocol is time-consuming than prepare-and-measure based QKD protocol as there is no fixed interval of time in which photon can arrive, unlike BB84 protocol. 

\section{\label{sec:conc}Conclusion}
In conclusion, we have demonstrated an entanglement based QKD that is BBM92 protocol over 35 m and 200 m free-space atmospheric channel, and simultaneously studied the effect of aerosols on the secure key rate.
This is the first study of its kind where the extinction coefficient of atmospheric aerosols is used to study the variation of entanglement, QBER and key rate. 
We also found that the key rate follows the same Beer-Lambert's law and the extinction coefficient of the atmospheric aerosols on that particular day.
As the key rate depends on the channel length, the larger the channel, the smaller will be the key rate. 
Further experiments are planned to study the effect for longer duration and to check the validity of models for estimating the key rate due to atmospheric conditions for the satellite based quantum communications \cite{VVM}. 
The presented results may find application in setting up large scale quantum communication network using satellites and the placement of entanglement photo-pair sources.

\section*{Acknowledgement}
Authors thank Dr. T. A. Rajesh and Vishnu Kumar Dhaker for providing the aerosol extinction and particulate matter data for 8 and 10 May 2021.  
Authors also acknowledge the financial support from DST through the QuST program.

\section*{Disclosures}
The authors declare no conflicts of interest related to this article.

\bibliography{manuscript}

\end{document}